\documentclass[trackchanges,twocolumn]{aastex7}

\usepackage[T1]{fontenc}
\usepackage{threeparttable}
\usepackage[normalem]{ulem}
\usepackage[version=4]{mhchem} 
\usepackage{float}
\usepackage{placeins}
\usepackage{capt-of}
\usepackage{footnote}

\begin{document}

\title{A 2 au resolution view by ALMA of the planet-hosting WISPIT 2 disk}

\author[orcid=0000-0003-4689-2684]{Facchini, Stefano}
\affiliation{Dipartimento di Fisica, Universit\`a degli Studi di Milano, Via Celoria 16, 20133 Milano, Italy}\email[show]{stefano.facchini@unimi.it}

\author[0000-0003-2045-2154]{Pietro Curone} 
\affiliation{Departamento de Astronom\'ia, Universidad de Chile, Camino El Observatorio 1515, Las Condes, Santiago, Chile}
\email[]{}

\author[0000-0000-0000-0000]{Myriam Benisty}
\affiliation{Max-Planck Institute for Astronomy (MPIA), Königstuhl 17, 69117 Heidelberg, Germany}
\email[]{}

\author[0000-0001-6417-7380]{Francesco Zagaria}
\affiliation{Max-Planck Institute for Astronomy (MPIA), Königstuhl 17, 69117 Heidelberg, Germany}
\email[]{}

\author[0000-0003-1534-5186]{Richard Teague}
\affiliation{Department of Earth, Atmospheric, and Planetary Sciences, Massachusetts Institute of Technology, Cambridge, MA 02139, USA}
\email[]{}

\author[0000-0001-7255-3251]{Gabriele Cugno}
\affiliation{Department of Astrophysics, University of Z\"urich, Winterthurerstrasse 190, 8057 Z\"urich, Switzerland}
\email[]{}

\author[0000-0001-7258-770X]{Jaehan Bae}
\affiliation{Department of Astronomy, University of Florida, Gainesville, FL 32611, USA}
\email[]{}

\begin{abstract}
We present deep, high spatial resolution interferometric observations of 0.88~mm continuum emission from the TYC 5709-354-1 system, hereafter WISPIT~2, obtained with the goal of detecting circumplanetary emission in the vicinity of the newly discovered WISPIT~2b planet. Observations with the most extended baseline configuration offered by ALMA, achieving an angular resolution of $25 \times 17$~mas ($3.3\times2.2$~au), revealed a single, narrow ring with a deprojected radius of $144.4$~au and width of $7.2$~au, and no evidence of circumplanetary emission within the cavity. Injection and recovery tests demonstrate that these observations can rule out point-like emission at the location of WISPIT~2b brighter than ${\approx}~45~\mu$Jy at the $3\sigma$ level. While these data can rule out PDS~70c like circumplanetary emission, the upper limit is consistent with empirical mass-flux relationships extrapolated from the stellar regime. Visibility modeling of the continuum ring confirms that WISPIT~2b lies significantly interior to the mm dust ring, raising doubts about the ability of WISPIT~2b to be the only driver of the dust structure. Possible solutions include either another lower mass companion, residing between WISPIT~2b and the cavity edge, likely in the gap seen by SPHERE at $\sim130\,$au, or that WISPIT~2b is either substantially more massive than IR-photometry based estimates (${\sim}~15~M_{\rm Jup}$) or on a moderately eccentric orbit. The combination of observations sensitive to the gas and dust distributions on larger spatial scales and dedicated hydrodynamical modeling will help differentiate between scenarios.
\end{abstract}



\section{Introduction}\label{sec:intro}
Planet–disk interactions are widely considered as a likely origin of the  substructures routinely revealed in high-resolution observations from ALMA in the sub-millimeter regime \citep[e.g.,][]{Andrews2020} and from infrared scattered-light instruments \citep[e.g.,][]{Benisty2023}. The most common substructures, rings and gaps, have often been attributed to unseen, embedded giant planets that carve dust-depleted gaps and create dust traps at the pressure maxima outside their orbits \citep[e.g.,][]{Pinilla2012}. While additional physical mechanisms (e.g., hydrodynamical instabilities) can produce annular substructures in disks, when applied to the population of observed rings and gaps, the planet interpretation would imply a large, yet-undetected population of giant planets at large separations \citep{Bae2023, Ruzza2025}. However, the search for protoplanets has been challenging, with many non-detections reported with direct imaging campaigns using both ground-based and space-based instruments \citep[e.g.,][]{Huelamo2022, Ren2023, Cugno2023}, possibly due to the circumstellar and circumplanetary disk material shielding the planet thermal emission \citep{Cugno2025}. 

Until recently, only PDS 70b and c, two giant planets detected within a common dust-depleted cavity surrounded by a double-peaked ring, provided a clear case of accreting protoplanets still embedded in a gas-rich disk \citep{Keppler_ea_2018, Haffert_ea_2019, Isella_ea_2019, Benisty2021}. However, recent  multi-epoch, multi-wavelength observations in different infrared (IR) bands ($H$, $K$ and $L'$) and in narrow H$\alpha$ band unveiled a new protoplanet, WISPIT~2b \citep{van_Capelleveen2025,Close2025}. This planet is embedded in a protoplanetary disk surrounding a $\sim1.08\,M_\odot$ star at a distance of 133\,pc \citep{van_Capelleveen2025}, and located at a separation of $0\farcs32$ (57~au from the astrometric fitting) from the central star. Assuming an age of 5.1$^{+1.3}_{-2.0}$\,Myr, the measured $H$ and $K_{\rm s}$ band infrared photometry by VLT/SPHERE led to an estimate of the planet mass of $4.9^{+0.6}_{-0.9}\,M_{\rm Jup}$ \citep{van_Capelleveen2025}. A consistent planet mass of $5.3\pm1.0\,M_{\rm Jup}$ was obtained from $L'$ band photometry with LBT/LMIRcam by \citet{Close2025}. Unlike the two PDS~70 planets, which orbit within a common wide cavity \citep{Keppler2019}, WISPIT~2b appears to reside within a gap, surrounded by two rings of small dust grains seen in scattered light. The detection in the H$\alpha$ filter by Magellan/MagAO-X \citep{Close2025} indicates that the planet is still accreting gas from its host protoplanetary disk at a rate of 2.25$^{+3.75}_{-0.17}\times 10^{-12}\,M_{\odot}$\,year$^{-1}$. 

Although the exact dynamical and physical structure of the gas in the immediate vicinity of protoplanets remains uncertain \citep[e.g.,][]{Lega2024}, it is generally thought that accretion onto a young planet is mediated by a circumplanetary disk \citep[CPD; e.g.,][]{aoyama18, Marleau2023}, which may also trap dust particles growing into satellitesimals \citep{BatyginMorby2020}. In the PDS 70 system, mm continuum emission has been detected near both planets \citep{Isella_ea_2019, Benisty2021, Fasano2025}, possibly indicating the presence of dust grains trapped in a CPD, although multi-wavelength observations of the material co-located with PDS 70c may indicate a contribution from non-thermal emission \citep{Dominguez2025}. The WISPIT~2b discovery now offers an additional laboratory to directly constrain planet-disk interactions, including dust trapping beyond the orbit of the planet, and search for any circumplanetary material in the vicinity of the accreting protoplanet.  

In this Letter, we examine the spatial distribution of dust continuum emission in the WISPIT~2 system using high-angular resolution ALMA observations. The paper is organized as follows: Sect.~\ref{sec:data} describes the observational setup and data reduction; Sect.~\ref{sec:results} presents the analysis of the dust continuum morphology and the search for CPDs; Sect.~\ref{sec:discussion} discusses our findings; and Sect.~\ref{sec:conclusions} provides a summary and our conclusions.

\section{Observations and Data Reduction}\label{sec:data}

WISPIT~2 was observed with ALMA in Band 7 in the most extended C-10 configuration as part of the DDT program 2024.A.00064.S (PI Facchini). The current dataset comprises three back-to-back Execution Blocks (EBs) obtained on 2025, September 10 for a total on-source integration time of 125~min. The wide hour-angle coverage results in excellent \emph{uv} sampling for this configuration. Additional observations providing intermediate and short baselines are expected by summer 2026. 

Band 7 was chosen as it offers the optimal trade-off between sensitivity and observing efficiency necessary to achieve the required detection significance. This choice was made assuming that the spectral index of the CPD around WISPIT~2b is comparable to that of PDS~70c, i.e., $\alpha = 2.0 \pm 0.2$ between Band 4 and Band 7 \citep{Dominguez2025, Fasano2025}. Higher and lower frequency Bands were excluded due to the correspondingly longer needed integration times.

The observations were carried out under excellent atmospheric conditions, with precipitable water vapor (PWV) between 0.25 and 0.4\,mm, leading to a high Execution Fraction (capped at 1.5 following standard ALMA procedures). Baselines span 132–15238\,m across 47 antennas, yielding a Maximum Recoverable Scale of $0\farcs27$ (substantially smaller than the $\gtrsim 2\arcsec$ diameter of the disk). Calibration was performed by the ESO ARC node, using J1912$-$0804 as phase calibrator and J1912$-$2914 for amplitude and bandpass calibration. The spectral setup consists of four Frequency Division Mode windows with 3840 channels each (no online averaging), providing a total continuum bandwidth of 7.5\,GHz. The central rest frequencies (LSRK) are 334.7, 336.6, 346.6, and 348.5\,GHz. The lowest-frequency window exhibits an anomalously high phase rms compared to the others, however dedicated tests show that this does not impact the image quality.

The substantial spatial filtering arising to the extended configuration prevented effective self-calibration. We tested conservative phase-only self-calibration using models derived either from a \texttt{CLEAN} image or from the best-fit parametric disk model (Section~\ref{sec:galario}); in both cases, the procedure failed to reduce imaging artifacts, since they are dominated by spatial filtering. Amplitude self-calibration was not attempted because the lack of short spacings compromises the derivation of reliable low spatial frequency amplitude gains. The upcoming shorter baselines data will allow this approach.

\begin{figure*}[ht]
    \centering    \includegraphics[width=\textwidth]{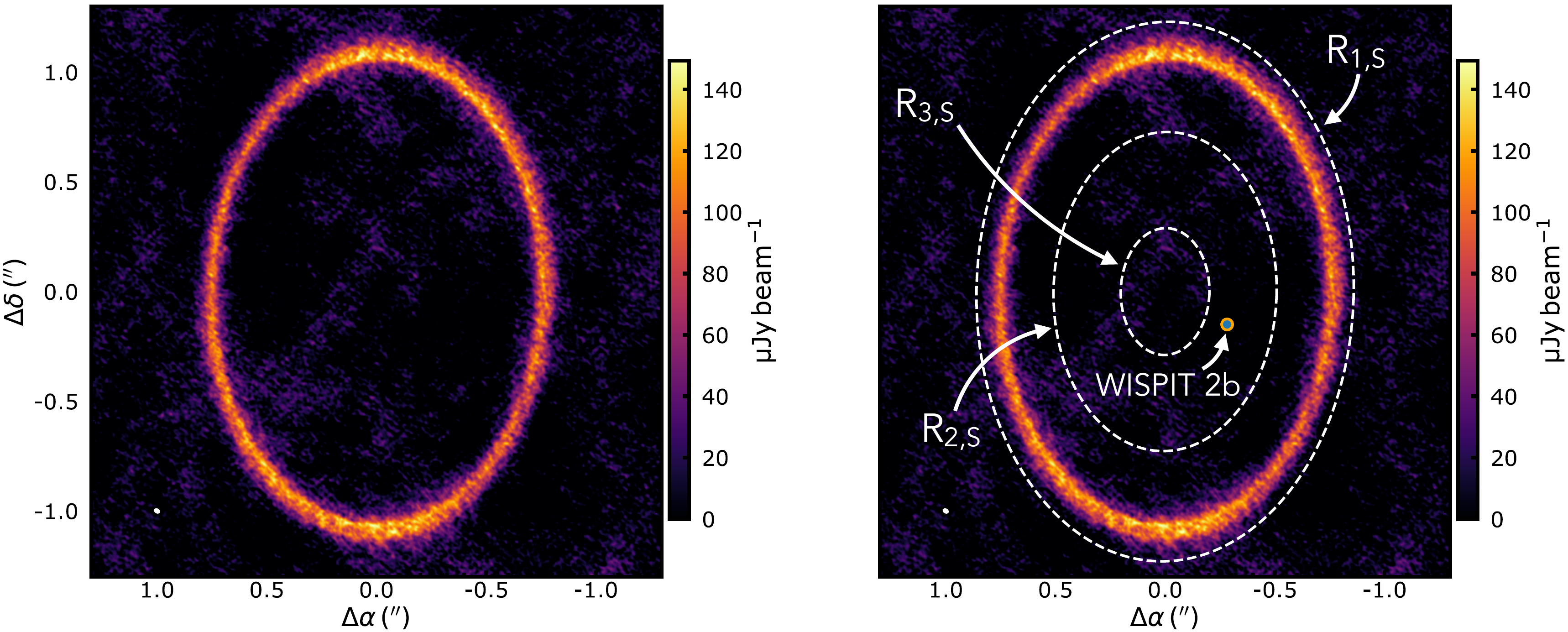}
    \caption{0.88\,mm continuum intensity of the WISPIT~2 system, showing a thin ring at $1\farcs086$  ($144.4\,$au deprojected distance). The right panel shows the same image, with white dashed lines indicating the radial peaks observed in IR scattered light by \citet{van_Capelleveen2025}, projected onto the disk midplane. The numbering ($R_{N,{\rm S}}$) follows the one by \citet{van_Capelleveen2025}. The location of the WISPIT~2b planet \citep{van_Capelleveen2025,Close2025} is highlighted in the right panel. The ALMA resolution element is shown in the bottom left of each panel.}
    \label{fig:WISPIT_2}
\end{figure*}

For imaging preparation, we first flagged channels associated with expected bright lines ($^{12}$CO $J$=3–2, SO $9_8$–$8_7$, C$_2$H $N$=4–3) and binned the remaining data into 62.5\,MHz channels. Continuum images were produced using Briggs weighting with a range of robust parameters and \emph{uv}-tapers, at a weighted frequency of 341.6\,GHz. All reconstructions show non-negligible artifacts induced by spatial filtering, including pronounced North–South ripples that elevate the rms within the central cavity of the ring-like structure. However, while the addition of shorter spacings will substantially improve imaging fidelity, the present dataset already enables robust scientific interpretation, in particular for point source detection.

Our fiducial image adopts a Briggs \texttt{robust} parameter of 1.0, pixel size of $0\farcs0025$ ($\approx 1/8$ of the beam minor axis), multiscale \texttt{CLEAN}ing with scales [0, 7, 21, 35] (with the maximum scale covering the width of the ring), and a threshold of $42\,\mu$Jy. Results are consistent across alternative imaging choices, including significant \emph{uv}-tapers aimed at enhancing sensitivity to diffuse emission. The \texttt{CLEAN} mask traces a projected circular region of radius $1\farcs5$, inclined by $44^\circ$ with a position angle of $359^\circ$ \citep{van_Capelleveen2025}. The rms measured within a projected circle of radius $0\farcs8$ is $16.6\,\mu$Jy, approximately $1.5\times$ above the expected thermal noise, indicating phase noise and imaging systematics. The synthesized beam of the fiducial image is $24\times17$\,mas ($3.3\times2.2$\,au), PA $=57.6^\circ$.

\section{Results}
\label{sec:results}
\begin{figure*}[ht]
    \centering    \includegraphics[width=\textwidth]{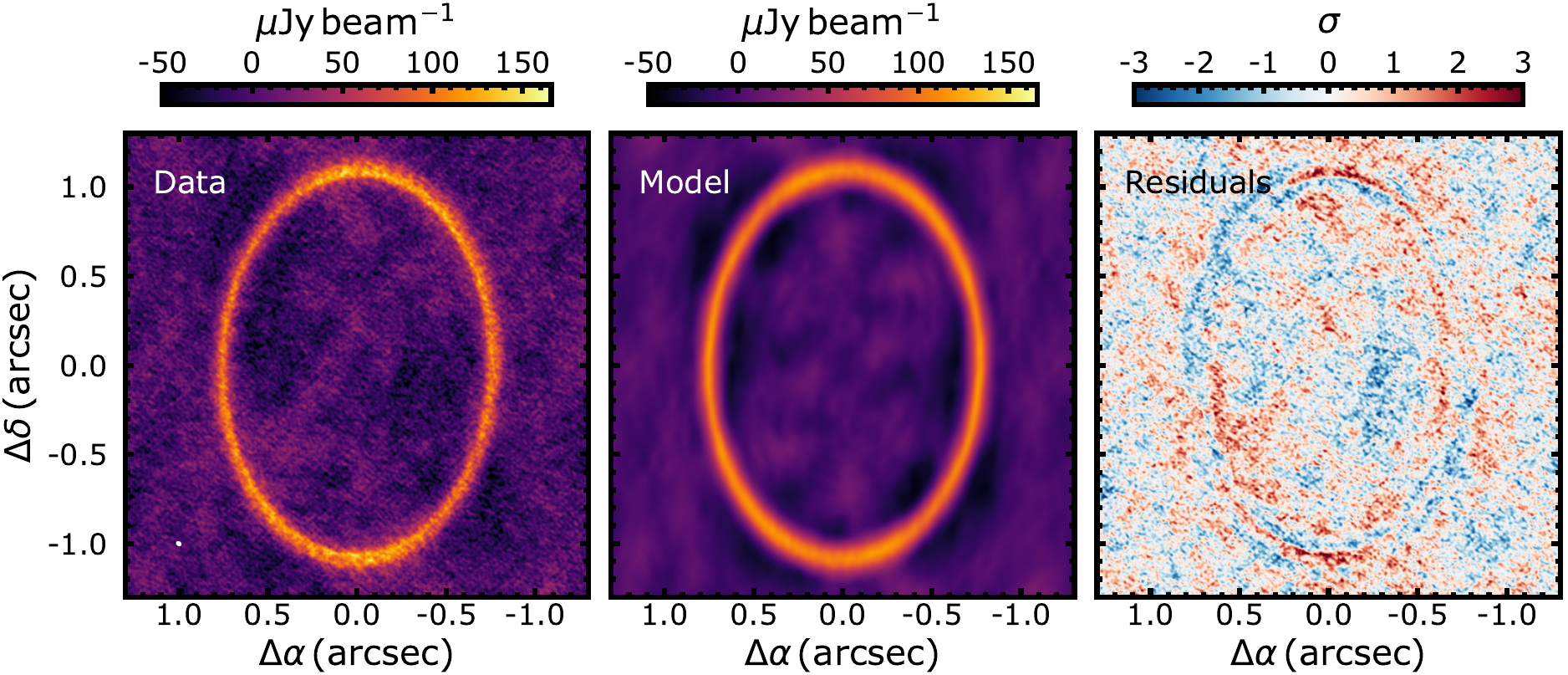}
    \caption{Data, best fit \texttt{galario} model, and residuals of WISPIT~2. Both the model and the residuals intensity maps have been reconstructed with the same \texttt{CLEAN}ing parameters as the original data. The rms ($\sigma=16.6\,\mu$Jy) used in the residuals colorbar is the one estimated within the mm-cavity on the data (see Section~\ref{sec:data}).}
    \label{fig:ring_imaging}
\end{figure*}

The ALMA image of WISPIT~2 (Figure~\ref{fig:WISPIT_2}) shows a prominent thin ring at a distance of $1\farcs086$. With the selection of baselines available, no additional radial structure is observable in these data. The mm-cavity is extremely wide, being one of the largest mm-cavities observed to date \citep[e.g.,][]{vanderMarel2023}. However, we caution the readers not to over-interpret the reconstructed image for three main reasons: 1) we are not sensitive to emission with spatial scales $\gtrsim0\farcs27$ (we are effectively imaging a high-pass-filter intensity map); 2) the extremely high angular resolution limits the sensitivity to low surface brightness emission; 3) imaging artifacts induce azimuthal variations in the emission of the mm ring. This third effect is particularly clear in the large negative emission at ${\rm PA}\sim45^\circ$ and $\sim225^\circ$ in the inner cavity (see Figure~\ref{fig:ring_imaging}). However, it is possible that large scale azimuthal modulations are present in the data, but are filtered out in the C-10 array configuration. No emission (thermal or not) is seen at the center of the disk, thus we cannot conclude where the true position of the star is.

\begin{deluxetable*}{CCCCCCC}
\tablecaption{Best fit parameters of the Gaussian-ring \texttt{galario} model of WISPIT~2. Uncertainties are computed from the 16th and 84th percentiles of the marginalized posterior distributions. The displacements of the center of the annulus are relative to the absolute ICRS coordinates $19^{\mathrm h}23^{\mathrm m}17.043704^{\mathrm s}$, $-07^\circ40'\,55.77275''$. \label{tab:galario}}
\tabletypesize{\footnotesize}
\tablehead{
\colhead{$I_0$}  & \colhead{$R_{\rm ring}$} & \colhead{$\sigma_R$} & \colhead{$i$} & \colhead{PA} & \colhead{$\Delta $RA} & \colhead{$\Delta$Dec} \\ 
\colhead{(Jy\,sr$^{-2}$)}  & \colhead{($\arcsec$)} & \colhead{($\arcsec$)} & \colhead{($^\circ$)} & \colhead{($^\circ$)} & \colhead{(mas)} & \colhead{(mas)} \\ 
}
\decimals
\startdata          
9.996^{+0.001}_{-0.001}              & 1.08613^{+0.00001}_{-0.00001}          & 0.0544^{+0.0001}_{-0.0001} & 45.6608^{+0.0005}_{-0.0003} & 178.6396^{+0.0004}_{-0.0005}     & -4.1^{+0.1}_{-0.1}    & -1.7^{+0.1}_{-0.1} 
\enddata
\end{deluxetable*}

\subsection{Parametric model of the sub-mm ring}\label{sec:galario}

In order to better reconstruct the emission of the WISPIT~2 ring given the limited $uv$ coverage, we model the visibility data with a single, circular, thin Gaussian ring, with the intensity described as:
\begin{equation}
    I(R) = I_0 \exp{\left(\frac{-(R-R_{\rm ring})^2}{2\sigma_R^2}\right)},
\end{equation}
with $R$ being the radius in cylindrical disk coordinates. The model is Fourier-transformed and sampled at the same {\it uv}-points as the observations using the \texttt{galario} code \citep{galario}. The fit includes $I_0$, $R_{\rm ring}$ and $\sigma_R$ as free parameters, together with the  geometrical parameters defining the disk center ($\Delta$RA, $\Delta$Dec), the disk inclination ($i$), and its position angle (PA). The best fit model is obtained by exploring the likelihood with the \texttt{emcee} package \citep{emcee}, with 100 walkers and 18000 steps, after a 8000 steps burn-in. The fit well converges to a radially resolved thin ring peaked at $1\farcs086$ (144.4\,au), with a width $\sigma_R = 0\farcs054$ (7.2\,au). The uncertainties and additional best fit values are reported in Table~\ref{tab:galario}, where uncertainties neglect the systematics associated to the inflexibility of the parametric model. The retrieved inclination and position angle are in agreement with the values obtained in $H$-band and $K_{\rm S}$-band \citep{van_Capelleveen2025}, which were also obtained under the assumption of circular morphology of the emission. The match in inclination and position angle between the IR and ALMA data indicates that comparable level of eccentricity would be present in the two datasets. Both the model and residual visibilities are imaged as the original data (Fig.~\ref{fig:ring_imaging}), with the residuals showing low significance, but structured residual patterns ($\lesssim 3\sigma$), indicating that a Gaussian representation is not perfect. The best-fit model yields a total disk flux density of $151.4\pm 0.3\,$mJy, where the uncertainty represents the 16th and 84th percentiles of the total flux density distribution derived from the last 1000 model iterations. The azimuthally averaged, deprojected visibilities, and the best-fit \texttt{galario} model are shown in Appendix~\ref{app:visib}.

\subsection{CPD injection-recovery}
We search for any unresolved emission at the location of the WISPIT~2b planet to identify potential thermal or non-thermal radiation associated with the planet or its circumplanetary disk. Several approaches have been used in the literature to isolate compact CPD emission \citep[e.g.,][]{Isella_ea_2019, Benisty2021, Andrews2021}. In our data, however, inspection of the residual images after subtracting the \texttt{galario} best-fit model reveals no significant emission at the position of WISPIT~2b, regardless of the chosen imaging parameters.

\begin{figure}[t]
    \centering    \includegraphics[width=\columnwidth]{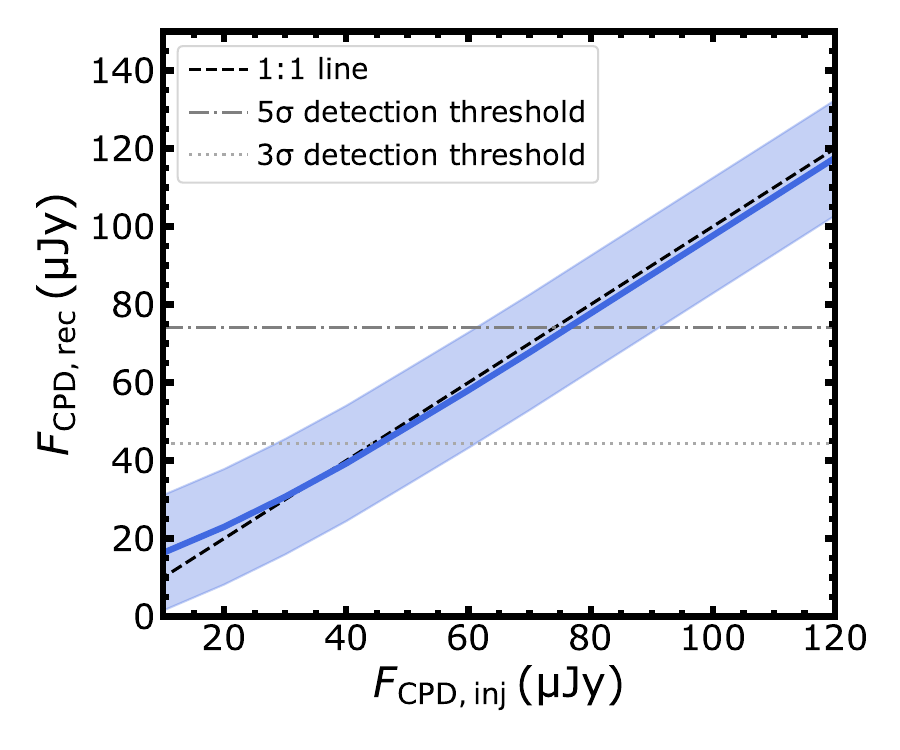}
    \caption{Injected ($F_{\rm CPD,inj}$) vs recovered ($F_{\rm CPD,rec}$) flux density of a point-source at the location of WISPIT~2b. The recovered flux density is estimated as the peak intensity within 1.5 beams at that location. The ribbon indicates the $1\sigma$ uncertainty on the peak intensity estimate.}
    \label{fig:CPD_flux_recovery}
\end{figure}

To derive an upper limit on the CPD flux density, we must account for imaging artifacts introduced by spatial filtering, manifesting as correlated noise and large-scale ripples across the reconstructed images. We therefore adopt an injection–recovery methodology following \citet{Andrews2021}. Specifically, we subtract the best-fit \texttt{galario} model from the visibilities and then inject a point source in visibility space at the planet location reported by \citet{van_Capelleveen2025} ($r=0\farcs320$, ${\rm PA}=242.1^\circ$). The injection is implemented by modifying the publicly available scripts of \citet{Andrews2021}. We then re-image the data using \texttt{CLEAN} parameters optimized to minimize the rms within the large cavity, which are the same as for the fiducial image. The recovered peak intensity is measured within an elliptical region centered on the injected source and with semi-axes 1.5 times the \texttt{CLEAN} beam.

We repeat this procedure for a number of CPD flux densities ranging from $10$ to $200\,\mu$Jy. Figure~\ref{fig:CPD_flux_recovery} shows the corresponding recovered flux densities, while in Appendix~\ref{app:injection_CC} we report images of the injection-recovery tests for CPD fluxes of 20, 50 and 80 $\mu$Jy. The uncertainty on the CPD flux density is taken as the rms in the cavity of the residual image prior to any injection, which is $14.8\,\mu$Jy (note that this is lower than the $16.6\,\mu$Jy computed on the full data). This yields a $5\sigma$ upper limit of $75\,\mu$Jy (or $45\,\mu$Jy at $3\sigma$) for a point-source CPD. This limit will improve as shorter baselines become available, enhancing image fidelity and point-source sensitivity. Finally, we emphasize that this constraint applies only to point-source emission; any spatially resolved CPD would exhibit lower surface brightness and could therefore remain undetected under these conditions (see Section~\ref{sec:disc_CPD}).

With the same approach, we also search for any point-source emission co-located with the companion candidate CC1 by \citet{Close2025}, at 15\,au deprojected distance from the central star ($r=0\farcs110$, ${\rm PA}=192^\circ$ in sky coordinates). No point-source is seen in the data. The results of the injection-recovery procedure for CC1 are shown in Appendix~\ref{app:injection_CC}.

\section{Discussion}
\label{sec:discussion}

\subsection{Interpreting the CPD detection limits}
\label{sec:disc_CPD}

Although a non-detection, the excellent sensitivity achieved with these observations allow us to place meaningful constraints on the properties of the CPD. If the emission is assumed to be optically thin, then the flux upper limit provides stringent upper limits on the total mass of the CPD, while for optically thick emission that same upper limit would yield a maximum radius of the CPD. In both cases, a CPD temperature, weighted by surface area, is needed. We use the same temperature as the one assumed for PDS~70c, i.e. 26\,K \citep{Benisty2021}, in order to allow a direct comparison with the only CPD detection in mm continuum known to date.  Under the optically thin assumption, we use two different opacities to estimate the upper limit on the CPD dust mass representing two scenarios for the dominant grain sizes \citep{Birnstiel2018}: 3.50\,cm$^{2}\,$g$^{-1}$ for 1\,mm-sized grains, and 0.76\,cm$^{2}\,$g$^{-1}$ for 1\,$\mu$m-sized grains. Using the full Planck-law, the 5$\sigma$ upper limit on the CPD flux density yields an upper limit of $0.0090\,M_\Earth$ ($0.73\,M_{\rm Moon}$) and $0.0416\,M_\Earth$ ($3.38\,M_{\rm Moon}$) assuming mm-sized or $\mu$m-sized grains, respectively.

Similarly, we can estimate an upper limit on the size of the CPD, in the assumption of optically thick emission over the full disk extent. Using the same temperature as for the optically thin case ($26\,$K), the upper limit on the CPD flux density relates to an upper limit on the mm radius of the CPD, with the following relation:
\begin{equation}
    R_{\rm CPD} = \sqrt{\frac{d^2 F_{\rm CPD}}{\pi B_\nu(T)\cos{i} }},
    \label{eq:R_CPD}
\end{equation}
where $d$ is the distance of the WISPIT~2 system \citep[133\,pc,][]{van_Capelleveen2025}, $B_{\nu}(T)$ is the Planck function, $F_{\rm CPD}$ the CPD flux density, and $i$ the inclination of the CPD, assumed to be the same as for the circumstellar disk. The $5\sigma$ upper limit on the CPD flux density provides an upper limit on $R_{\rm CPD}$ of $0.62\,$au. This is smaller than the expected size of the CPD from a dynamical perspective, from which the size in gas of a CPD can be estimated as one third of the Hill radius \citep[e.g.,][]{ayliffe2009}. By using a stellar mass of $1.08^{+0.06}_{-0.17}M_\odot$, and a mass for WISPIT~2b of $4.9^{+0.9}_{-0.6}M_{\rm Jup}$ \citep{van_Capelleveen2025}, a Monte Carlo sampling of the two masses with asymmetric Gaussian distributions yields an expected radius for the CPD in gas of $R_{\rm gas,exp} = 2.07^{+0.13}_{-0.12}\,$au. We can therefore exclude the presence of an optically thick CPD, with an extent of the continuum emission that is as large as one third of the Hill radius. The same result was obtained by \citet{Isella_ea_2019} for PDS~70c.

Given the exceptionally high angular resolution of our ALMA observations, a CPD with a radius equal to one third of the Hill radius would be spatially resolved in our fiducial images, causing its emission to be distributed over multiple resolution elements. For completeness, we thus calculated the rms within the cavity of the residuals using progressively larger \texttt{CLEAN} beams in the image reconstruction. The corresponding rms scaling is presented in Appendix~\ref{app:injection_CC}. 

\subsection{Comparing the mm flux density of WISPIT~2b with other systems}

\begin{figure}[t]
    \centering    \includegraphics[width=\columnwidth]{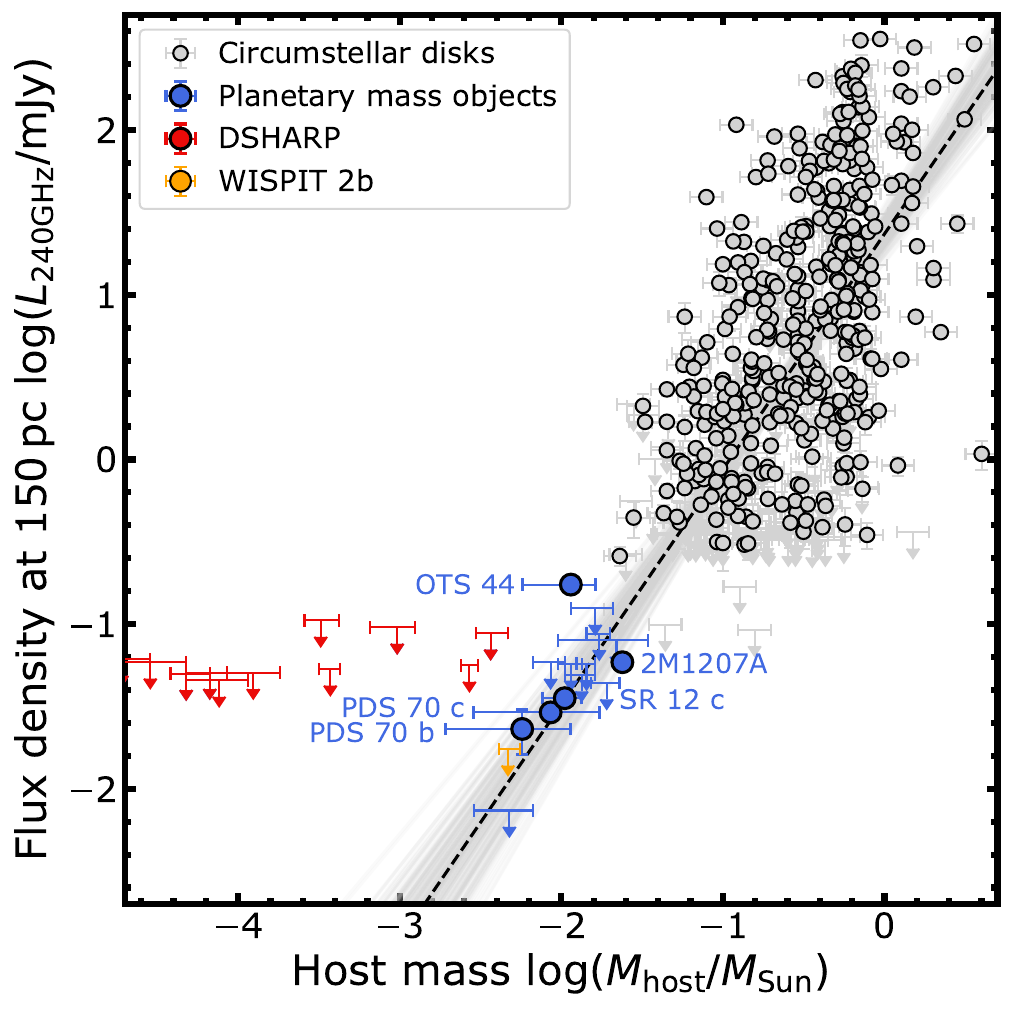}
    \caption{Circumstellar and circumplanetary flux density (rescaled to $240\,{\rm GHz}$ and $150\,{\rm pc}$) distribution as function of their host mass. Most planetary mass objects closely follow the trend (best-fit linear correlation, black dashed line) expected for circumstellar disks.}
    \label{fig:CPD_scaling}
\end{figure}

In Figure~\ref{fig:CPD_scaling} we place our ($3\sigma$) flux density upper limit in an empirical context by comparing it with the distribution of circumstellar and circumplanetary (sub)mm flux density versus host mass (as similarly shown by \citealt{Wu2020,Andrews2021}). Flux densities have been rescaled to a common frequency of $240\,{\rm GHz}$ with a spectral index $\alpha=2.2$ for circumstellar disks \citep[taken as the average of the $0.9$ - $1.3$ mm spectral indices of circumstellar disks,][]{Andrews2020} and 2.0 for circumplanetary emission \citep[using the case of PDS~70c as the only available benchmark,][]{Fasano2025,Dominguez2025}, and to a distance of $150\,{\rm pc}$. Dots and down-pointing arrows indicate mm-continuum detections and $3\sigma$ upper limits, respectively. Circumstellar disks in low-mass nearby star formation regions (from the catalog of \citealt{Manara2023}) are shown as gray dots. The black dashed line indicates their best-fit linear correlation (with slope $\beta=1.43\pm0.10$ and intercept $\alpha=1.37\pm0.06$) obtained using the linear regression method implemented in \texttt{linmix} \citep{Kelly2007}; single chain realizations are also shown (as lighter gray lines) to give a sense of the scatter around the best-fit curve. Over-plotted in blue are the results of similar searches for (sub)mm continuum emission around low/planetary-mass companions in 
GSC~6214-210~B \citep{Bowler2015}, GQ~Lup \citep{MacGregor2017}, 2MASS~J12073346-3932539~A and B (2M1207, \citealt{Ricci2017}), SR~12~c \citep{Wu2022}, the sample of \citet{Wu2020}, the isolated planetary mass object OTS~44 \citep{Bayo2017}, and the planets PDS 70~c \citep{Benisty2021,Fasano2025} and b (although potentially still contaminated by inner disk emission, \citealt{Fasano2025}). The masses of PDS~70b and c were obtained from astrometric constraints by \citet{Trevascus2025}, where we considered the large uncertainties they provide to be compatible with photometry-based masses \citep{Wang2020}. Finally, upper limits on the circumplanetary emission around the protoplanets proposed to be carving (some of) the DSHARP gaps, defined as the flux densities corresponding to a recovery fraction of at least 50\% \citep{Andrews2021}, are also displayed in red (for the most-recent planet mass estimates of \citealt{Ruzza2025}). Our $3\sigma$ upper limit on (sub-)mm continuum emission towards WISPIT~2b is shown as a yellow arrow. {As noted by \citet{Wu2020,Andrews2021}, low mass objects in the Brown Dwarf and planetary regime have flux densities well in line with those expected from a blind extrapolation of the correlation in place among circumstellar disks. The non-detection of a mm point-source co-located with WISPIT~2b is consistent with this empirical correlation. However, the large scatter around the best-fit relation for circumstellar disks, combined with both the low flux densities expected from circumplanetary emission and the small number of detections, makes it extremely challenging to determine whether a different correlation applies to circumplanetary disks.

It is also meaningful to directly compare these results with the mm flux density measured for PDS~70c, in the assumption that scalings of flux densities of CPDs deviate from the trend found for stellar and Brown Dwarf mass objects, since the formation pathways of their disks is expected to be very different. Under the assumption that a CPD flux density linearly scales with the Hill radius of the planet, we would expect, following Equation~\ref{eq:R_CPD}:
\begin{equation}
F_{\rm CPD} \propto \left( \frac{a_{\rm p}}{d} \right)^2 \left( \frac{M_{\rm p}}{M_*} \right)^{2/3}B_{\nu}(T)\cos{i},
\end{equation}
with $a_{\rm p}$ and $M_{\rm p}$ being the orbital radius and mass of the planet, respectively. Adopting a Band 7 flux density for PDS~70c of $111\pm12\,\mu$Jy, computed as the weighted average of the measurements reported by \citet{Fasano2025}, assuming the same CPD temperature for the two systems, and noting that the two circumstellar disks have similar inclinations, we obtain:
\begin{equation}
\frac{F_{\rm WISPIT~2b}}{200\,\mu{\rm Jy}} \sim \left(\frac{M_{\rm WISPIT~2b}}{M_{\rm PDS~70c}}\right)^{2/3},
\end{equation}
which suggests that, given the similar estimated planet masses in the two systems, mm emission co-located with WISPIT~2b should already be detectable. More detailed modeling, beyond the scope of this Letter, is required to assess whether physically motivated temperatures across the range of plausible planetary masses (for which systematic uncertainties are substantial in both cases) would modify this conclusion.

Under the optically thin assumption, for which the CPD flux density scales linearly with CPD dust mass, our $5\sigma$ upper limit of 0.009$\,M_\Earth$ on the CPD of WISPIT~2b (assuming 1\,mm grains) is comparable to the inferred CPD mass of PDS 70c, 0.008$\,M_\Earth$, derived using the same methodology \citep[rescaling the value by][to account for the different flux density of PDS~70c used in this Letter]{Benisty2021}. The two planets exhibit similar accretion rates of $\dot{M}_{\rm p}\sim2\times10^{-12}\,M_{\odot}\,$yr$^{-1}$ \citep{Close2025,Close2025_PDS70}. However, given the large systematic uncertainties in accretion rate estimates and the intrinsic variability of the H$\alpha$ line used to determine the accretion on the planets \citep[e.g.,][]{Close2025_PDS70,Zhou2025}, we cannot assess whether a significant difference exists in the CPD accretion timescale $t_{\rm acc}=M_{\rm CPD}/\dot{M}_{\rm p}$. At present, WISPIT~2b remains consistent with the estimates for PDS~70c, with $t_{\rm acc}\sim 1.2\,$Myr when assuming a gas-to-dust ratio of 100. 

\subsection{Comparison with SPHERE Data}

The single narrow ring observed in the mm continuum is in stark contrast to the gap-punctuated morphology traced by polarized scattered light \citep{van_Capelleveen2025}. Intriguingly, four concentric rings in the scattered light were observed between 38 and 316~au, while only a single ring in the mm continuum at 144~au was found, as shown in Figure~\ref{fig:WISPIT_2}. A clear result from this simple comparison is that large grains (as traced) by ALMA are predominantly confined at large orbital separation, due to particle trapping in a local pressure maximum. Smaller grains, as traced by IR scattered light from $H$ to $L'$ bands \citep{van_Capelleveen2025,Close2025}, must be well-dynamically coupled to the gas, and can be advected radially \citep[see discussion by][for the PDS~70 system]{Bae2019,Pinilla2024}. This configuration clearly shows that, not surprisingly, massive embedded planets are able to trap large particles in confined regions, and lead to the formation of wide cavities at mm wavelengths, as predicted by hydrodynamical modeling.

\subsection{An additional planet beyond WISPIT~2b?}\label{sec:another_planet}

One of the most striking features is the position of WISPIT~2b relative to the mm-cavity edge. Theoretical models have shown that the width of the gap opened by a planet (assuming a circular orbit) depends on its mass \citep[e.g.,][]{Kanagawa2016,Zhang2018}. However, for WISPIT~2b the distance to the cavity edge, $\sim90$\,au, would require a planet mass that is larger than inferred from infrared-photometry. To illustrate this point, \autoref{fig:R_mm_a_planet} displays the mm-continuum gap width normalized to the gap location as a function of the planet-to-star mass ratio for all detected sources (PDS~70b and c and WISPIT~2b, shown in color) and putative planets proposed to explain the observed continuum structure (shown as gray points). When a planet is detected, as is the case for PDS~70b and c and WISPIT~2b, the orbital radius of the planet ($a_{\rm p}$) is used, rather than the gap radius (see Caption of Figure~\ref{fig:R_mm_a_planet}). Overlaid are two analytical relationships obtained from hydrodynamical simulations by \citet{Rosotti2016} and \citet{Facchini2018} in two different planet mass regimes, but which show consistent scalings with planet masses.

\begin{figure}[t]
    \centering
    \includegraphics[width=\columnwidth]{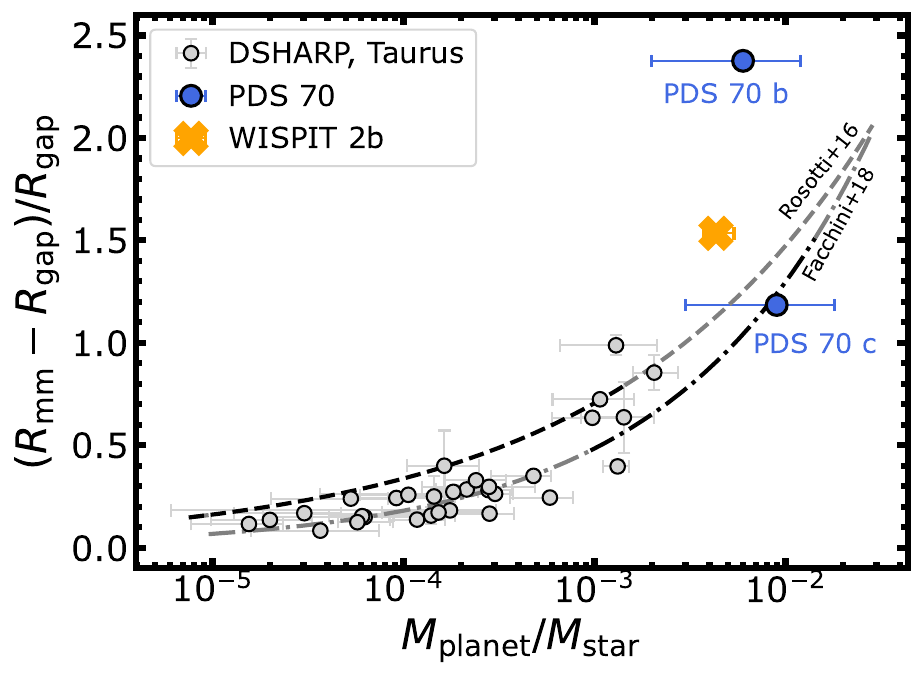}
    \caption{Gap width (in units of the gap radius) as a function of the planet-to-star mass ratio. For systems hosting detected protoplanets, $R_{\rm gap} = a_{\rm p}$, where $a_{\rm p}$ is the orbital radius of the planet. The candidate protoplanets in (some of) the gaps in the DSHARP sample \citep{Andrews_ea_2018,Huang2018} and the Taurus survey of \citet{Long2018} are displayed as gray dots, with putative planet masses estimated by \citet{Ruzza2025} and stellar masses by \citet{Andrews_ea_2018,Long2019}. PDS~70b and c are over-plotted in blue and WISPIT~2b in yellow. The dashed and dashed dotted black lines are the expected gap width to planet mass trends predicted by \citet{Rosotti2016} and \citet{Facchini2018}, their extrapolation to the whole planet mass range we considered is shown in gray.}
    \label{fig:R_mm_a_planet}
\end{figure}

For a majority of cases, including PDS~70c, the gap size normalized to the gap location closely follows the expected semi-analytical trend. This should not surprise with the non-detections, since the putative protoplanet masses were estimated from a grid of hydrodynamical simulations similar to the ones used to extract the analytical trends shown in Figure~\ref{fig:R_mm_a_planet}. PDS~70b is a clear outlier. Its mass is too low to explain the very wide (sub)mm cavity it was detected in, which led to the prediction of another massive companion within the cavity \citep{Keppler2019}, later confirmed by the detection of PDS~70c \citep{Haffert_ea_2019}. PDS~70c perfectly matches the expected gap width for its mass. We speculate WISPIT~2b may be in a similar case as PDS~70b: the planet is located too far from our newly detected (sub)mm ring for its estimated mass of $4.9\,M_{\rm Jup}$.

This discrepancy could be explained in four ways. The most tantalizing one is that, as in the case of PDS~70b, another companion is present between WISPIT~2b and the (sub)mm ring. The natural location for this planet to reside is the gap in scattered light detected by \citet{van_Capelleveen2025} at about $130\,{\rm au}$, yielding a $(R_{\rm mm}-R_{\rm {gap}})/R_{\rm gap}\sim0.1$. Assuming this semi-major axis, a sub-Jupiter mass companion follows the relationship predicted by the numerical simulations of \citet{Facchini2018}, and generates a mm ring at $144\,$au. If located at the inner edge of the scattered-light gap, at about $122\,{\rm au}$, such a companion would be in $3:1$ Mean Motion Resonance (MMR) with WISPIT~2b, although simulations will be needed to assess the dynamical stability of the system  since this is a second-order resonance. 

Alternatively, WISPIT~2b could be on an eccentric orbit, a hypothesis that needs to be re-assessed in the light of better orbital characterization of the companion. Current astrometric constraints by \citet{van_Capelleveen2025} indicate that the eccentricity of the planet's orbit should be $\lesssim0.3$. For a moderately large eccentricity of $0.2 - 0.3$, numerical simulations show that the outer pressure peak can be located $10-20 \%$ further away from the planet than that around a planet on a circular orbit (private communication; Jaehan Bae).
Thus, the high end of the allowed eccentricity range may be able to reconcile the size of the mm cavity and the planet masses estimated from IR photometry without an additional outer planet. Analysis of the gas kinematics will be able to reveal dynamical perturbations associated with eccentric planetary orbits \citep[e.g.,][]{Baruteau2021}.

Another option is that WISPIT~2b could be more massive than inferred by \citet{van_Capelleveen2025} and \citet{Close2025} from $H$ and $L'$-band photometry, respectively, from 10 up to $15\,M_{\rm Jup}$, to be consistent with the wide (sub)mm-continuum cavity. In this case, the gap in scattered light external to WISPIT~2b could be due to shadowing from the ring immediately inwards of it, due to a puffed-up gap edge, as seen in three-dimensional hydrodynamical simulations \citep[e.g.,][]{Bi2021}. A $15\,M_{\rm Jup}$ planet would also perfectly match the prediction of the ratio between $R_{\rm mm}$ and the IR wall ($R_{\rm wall}$) outside the planet orbit as seen in scattered light. From Figure~18 by \citet{van_Capelleveen2025}, we can estimate the radius of the IR wall just outside the orbit of WISPIT~2b to be $\sim80\,$au. Inverting equation~1 by \citet{dejuan_ovelar2013}, which presents the scaling of the radial separation of $R_{\rm wall}$ and $R_{\rm mm}$ with planet mass from a set of hydrodynamical simulations, a $R_{\rm wall}/R_{\rm mm}\sim0.55$ ratio would require a $15\,M_{\rm Jup}$ planet. However, the width of the gap in scattered light where WISPIT~2b resides would be too narrow to host such a massive planet, according to the predictions of \citet{Kanagawa2016}, and would require extremely high local diffusivities ($\alpha\approx$ a few $10^{-3}$ up to $10^{-2}$) to be consistent with those of \citet{Zhang2018}. Moreover, in order to reconcile the high planet mass with the IR photometry, we would need to invoke either some extinction along the line of sight to the planet, or a cold-start scenario for its evolutionary path. However, such a high mass planet would induce strong eccentricity in the disk gas. These strong perturbations are not seen in the well-coupled small grains traced by the IR scattered-light images by \citet{van_Capelleveen2025}. An analysis of the gas kinematics will promptly reveal any strong eccentric motions in the disk \citep[e.g.,][]{Kuo2022,Ragusa2024,Cuello2024}, and therefore confirm or reject this hypothesis. 

Lastly, our scalings of $(R_{\rm mm}-a_{\rm p})/a_{\rm p}$, and $R_{\rm wall}/R_{\rm mm}$, may be biased by the use of two-dimensional, locally-isothermal hydrodynamical simulations. A full three-dimensional treatment, with a less simplistic approach for the disk thermodynamics, may present slightly different scaling laws. However, \citet{Cordwell2025} demonstrated that the observed $(R_{\rm mm}-R_{\rm gap})/R_{\rm gap}$ values in disks require planets more massive than those inferred from 2D simulations, and therefore do not alleviate this tension.   Dedicated hydrodynamical simulations of the WISPIT~2 system are needed to fully assess this.

As a final note, we stress that the above conclusions are, to some extent, still dependent on the currently available stellar mass estimates. If in favor of a more massive central star, future accurate  estimates (e.g., from disk dynamics) could hasten the discrepancy between WISPIT~2b's mass and the gap width trends displayed in Figure~\ref{fig:R_mm_a_planet}, thus strengthening our hypothesis that an external companion might be present in the system.

\section{Conclusions}\label{sec:conclusions}

In this paper, we analyzed new high resolution ALMA data of the WISPIT~2 system, the second disk known to unambiguously host an embedded protoplanet \citep{van_Capelleveen2025,Close2025}, obtained to search for evidence of circumplanetary emission and to map the large grain distribution, offering a midplane comparison to the surface structure traced in scattered light. The conclusions of this study are the following:
\begin{itemize}
    \item The mm thermal emission shows a single, radially resolved, thin ($0\farcs054$ or 7.2au) ring peaking at $1\farcs086$ (144.4\,au), far beyond the location of WISPIT~2b (57~au).
    \item The radial separation of WISPIT~2b and the mm continuum ring is substantially larger than numerical models predict for a $5~M_{\rm Jup}$ planet. Assuming the photometrically-derived mass is robust, this implies that a second, lower mass planet lies between WISPIT~2b and the mm ring, analogous to the PDS~70 system. A natural location of this additional outer planet is the gap detected in scattered light at $\sim130\,$au. Alternatively, WISPIT~2b may occupy a moderately eccentric orbit.
    \item Another possibility is that WISPIT~2b is more massive than estimated by IR photometry. Reconciling such a higher mass ($>10\,M_{\rm Jup}$) would require some level of extinction toward the planet. In this scenario, gas kinematic observations should reveal clear signatures of orbital eccentricity.
    \item No mm emission is detected within the cavity. Injection-and-recovery tests demonstrate that point-source emission at the location of WISPIT~2b above ${\sim}~45~\mu$Jy can be ruled out at the $3\sigma$ level. This excludes that a potential CPD for WISPIT~2b shows optically thick continuum emission within one third of the Hill radius of the planet.
    \item The non-detection of a CPD at the location of WISPIT~2b is consistent with the extrapolation of empirical relationships of mm flux densities against host mass from the stellar mass regime. Assuming similar temperatures for PDS~70c and WISPIT~2b, we can however exclude that the two CPDs follow the same scaling with the planet Hill radius.
\end{itemize}

Taken together, these results highlight the power of combining millimeter continuum imaging with complementary shorter-wavelength observations in systems that host embedded protoplanets. Such multi-wavelength studies uniquely probe both the disk midplane and surface layers, enabling stringent tests of theoretical models of planet–disk interactions, gap opening, and dust trapping. They also provide crucial empirical constraints on the conditions under which circumplanetary disks may form, and thereby offer an essential pathway toward linking observed protoplanets to their expected formation and accretion environments. Upcoming and future observations of the gas dynamics and hydrodynamical models of WISPIT~2 will further constrain and benchmark the co-evolution of protoplanets and their host protoplanetary disk. 

\begin{acknowledgments}
The authors thank the ALMA Director for awarding time to this DDT program, and the referee for their constructive comments. They are also grateful to Alessandro Ruzza and Andrew Winter for insightful discussions. This paper makes use of the following ALMA data: 

ADS/JAO.ALMA 2024.A.00064.S.

ALMA is a partnership of ESO (representing its member states), NSF (USA) and NINS (Japan), together with NRC (Canada), MOST and ASIAA (Taiwan), and KASI (Republic of Korea), in cooperation with the Republic of Chile. The Joint ALMA Observatory is operated by ESO, AUI/NRAO and NAOJ. S.F. acknowledges financial contributions from the European Union (ERC, UNVEIL, 101076613) and from PRIN-MUR 2022YP5ACE. P.C. acknowledges support by the ANID BASAL project FB210003. M.B. has received funding from the European Research Council (ERC) under the European Union’s Horizon 2020 research and innovation programme (PROTOPLANETS, grant agreement No. 101002188). Views and opinions expressed are however those of the authors only and do not necessarily reflect those of the European Union or the European Research Council. Neither the European Union nor the granting authority can be held responsible for them. This work has been carried out within the framework of the NCCR PlanetS supported by the Swiss National Science Foundation under grant 51NF40\_205606. This work made use of the Geryon cluster at the Centro de Astro-Ingenieria UC, which is supported by ANID BASAL project FB21000, BASAL CATA PFB-06, Anillo ACT-86, FONDEQUIP AIC-57, and QUIMAL 130008.

\end{acknowledgments}





%
\facilities{ALMA}

\software{\texttt{CASA} v6.6.4 \citep{CASA2022},  
          \texttt{numpy} \citep{harris2020}, 
          \texttt{matplotlib} \citep{hunter2007},
          \texttt{galario} \citep{galario},
          \texttt{emcee} \citep{emcee}
          }

\bibliography{bibliography}{}
\bibliographystyle{aasjournalv7}

\appendix

\section{Azimuthally averaged visibilities of WISPIT~2}
\label{app:visib}

Figure~\ref{fig:uv_baselines} shows the azimuthally averaged visibilities, after deprojection and re-centering, with the best-fit \texttt{galario} model overplotted to the data. The marginalized posterior distributions of the \texttt{galario} fitting procedure are shown in Figure~\ref{fig:corner}.

\begin{figure}[h]
    \centering    \includegraphics[width=0.5\columnwidth]{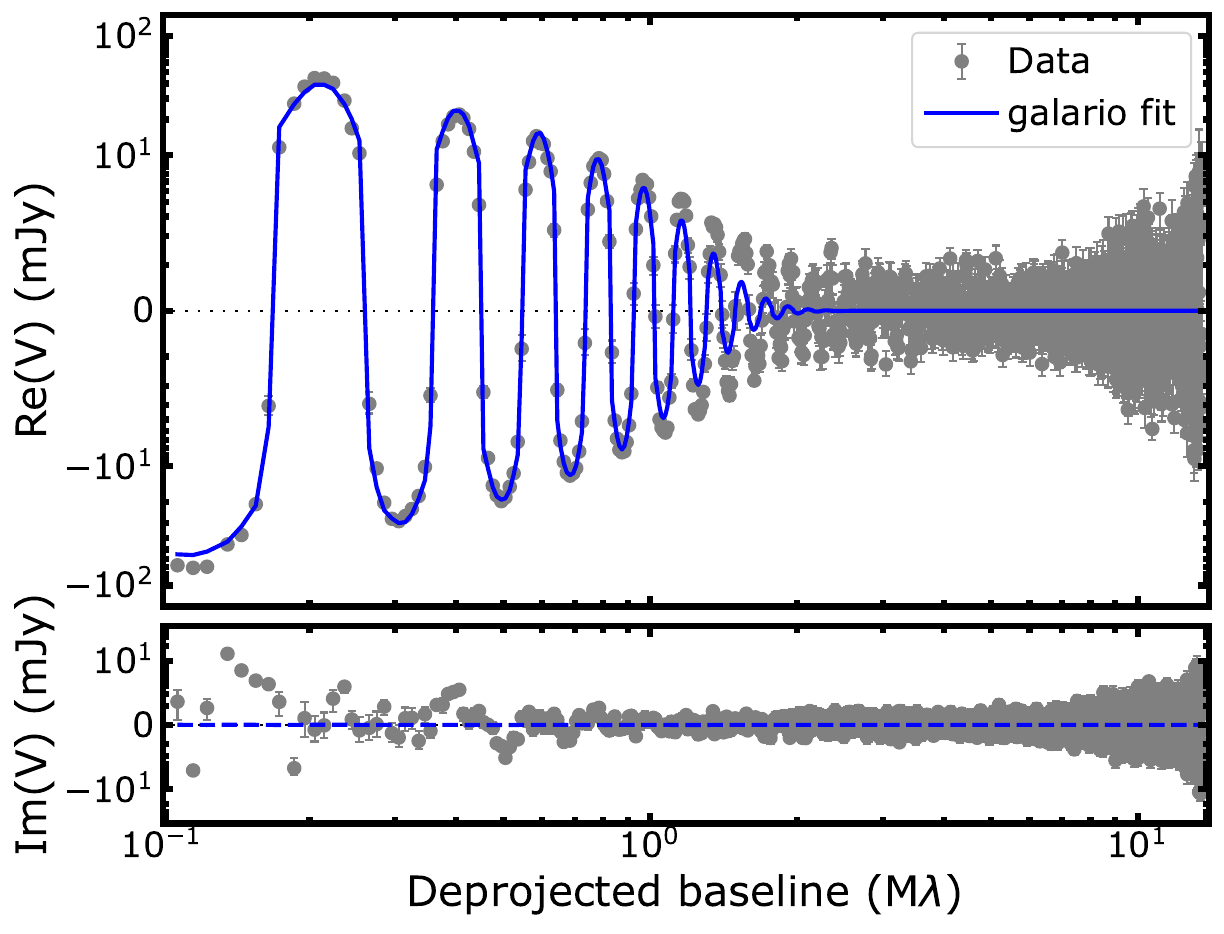}
    \caption{Azimuthal average of the Real and Imaginary parts of the visibilities after deprojection and re-centering, using the best fit geometrical parameters from the \texttt{galario} model, which is shown in blue. Spatial filtering is clear given the large shortest baseline available with these data.}
    \label{fig:uv_baselines}
\end{figure}

\begin{figure*}[ht]
    \centering    \includegraphics[width=0.8\textwidth]{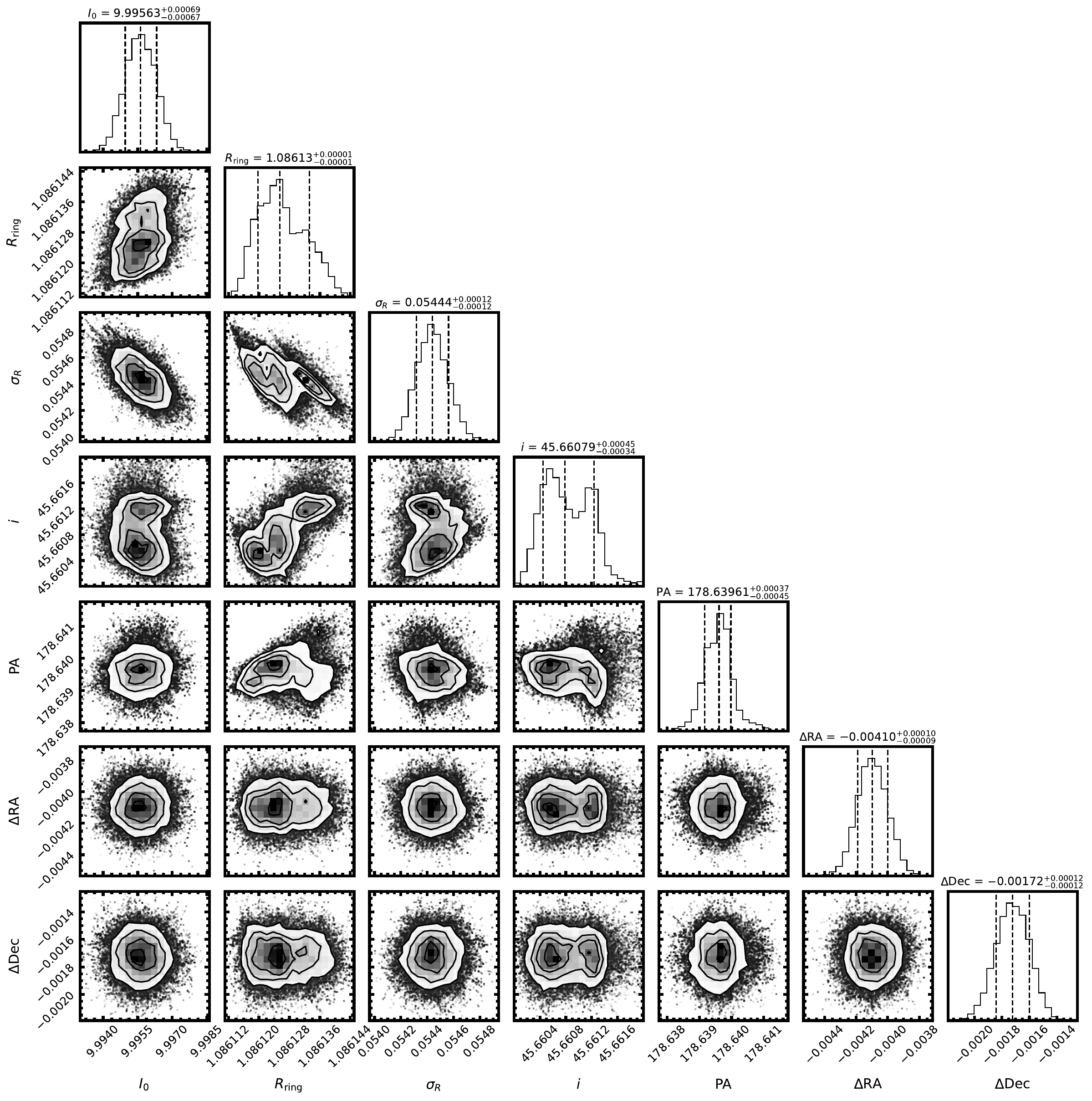}
    \caption{Marginalized posterior distribution of the seven parameters identifying the parametric \texttt{galario} model.}
    \label{fig:corner}
\end{figure*}

\section{Injection-recovery of a point-source at the location of WISPIT~2b and CC1}
\label{app:injection_CC}
We show reconstructed images of a CPD injected at the location of WISPIT~2b, using our default imaging parameters, in Figure~\ref{fig:CPD_imaging}, for a range of injected flux densities. In Figure~\ref{fig:CPD_resolution} we reconstruct the scaling of the rms within the WISPIT~2 mm cavity as a function of the resulting \texttt{CLEAN} beam, when progressively increasing the angular scale of $uv$-tapering in the imaging. 

Finally, we report the results of the injection-recovery of a point source at the location of CC1 in Figure~\ref{fig:CPD_flux_recovery_CC1}.

\begin{figure*}[ht]
    \centering    \includegraphics[width=\textwidth]{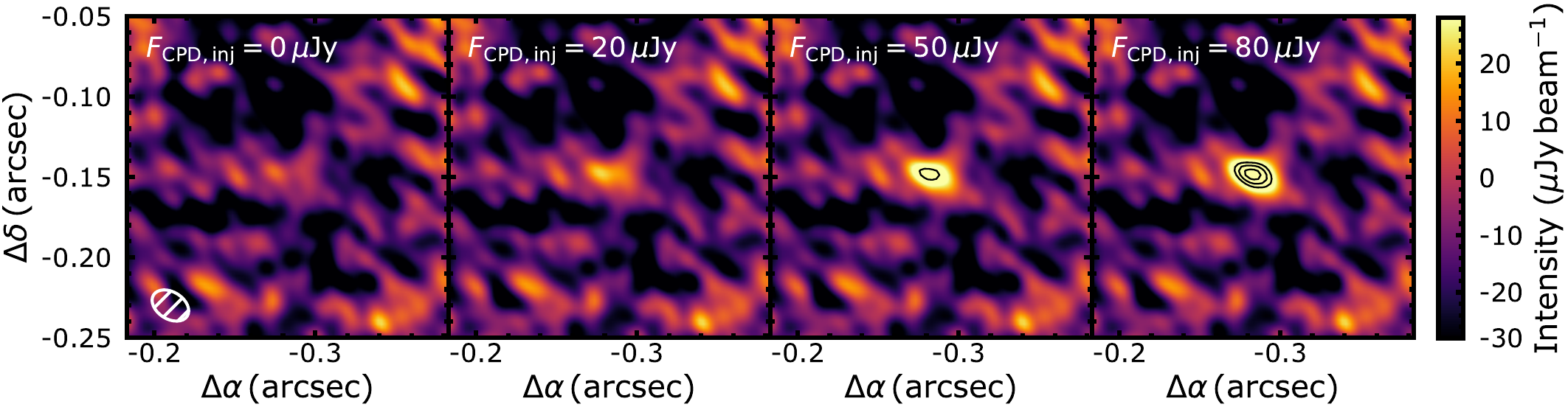}
    \caption{Examples of reconstructed images of a point source injected at the location of WISPIT~2b, located at the center of each frame, after subtracting the best fit \texttt{galario} model in the outer regions (not visibile on these angular scales). The coordinates are relative to the center of the ALMA ring. Black contours indicate the $[3,4,5]\sigma$ levels. The \texttt{CLEAN} beam is shown in the bottom right of the first panel.}
    \label{fig:CPD_imaging}
\end{figure*}

\begin{figure}[ht]
    \centering    \includegraphics[width=0.6\columnwidth]{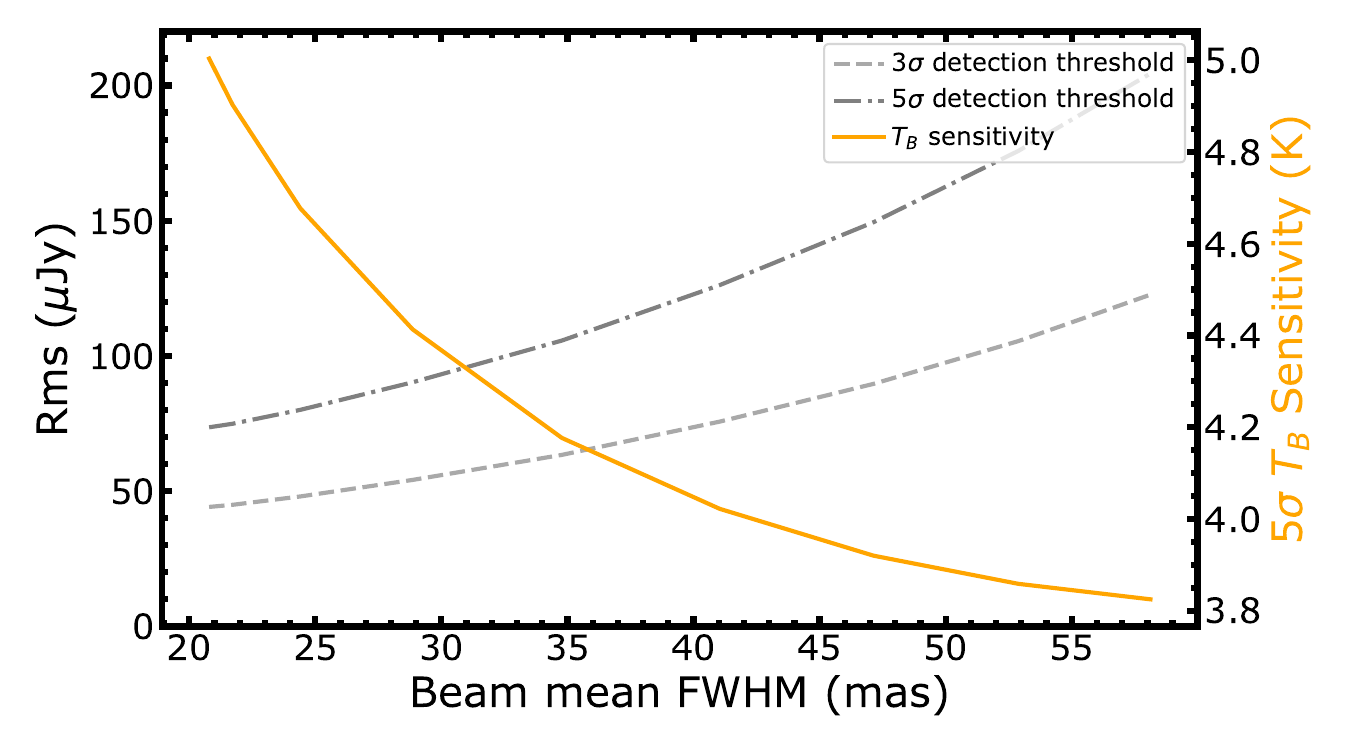}
    \caption{Rms estimated within the cavity of WISPIT~2b after removing the best-fit \texttt{galario} model, for a large set of imaging parameters, starting from Briggs weighting with \texttt{robust}=1.0, and progressively increasing additional $uv$-tapering ($[5,10,15,20,25,30,35,40]$~mas). The orange line (portrayed against the $y$-axis on the right side of the panel) shows the $5\sigma$ brightness temperature sensitivity using the full Planck law.}
    \label{fig:CPD_resolution}
\end{figure}

\begin{figure}[ht]
    \centering    \includegraphics[width=0.5\columnwidth]{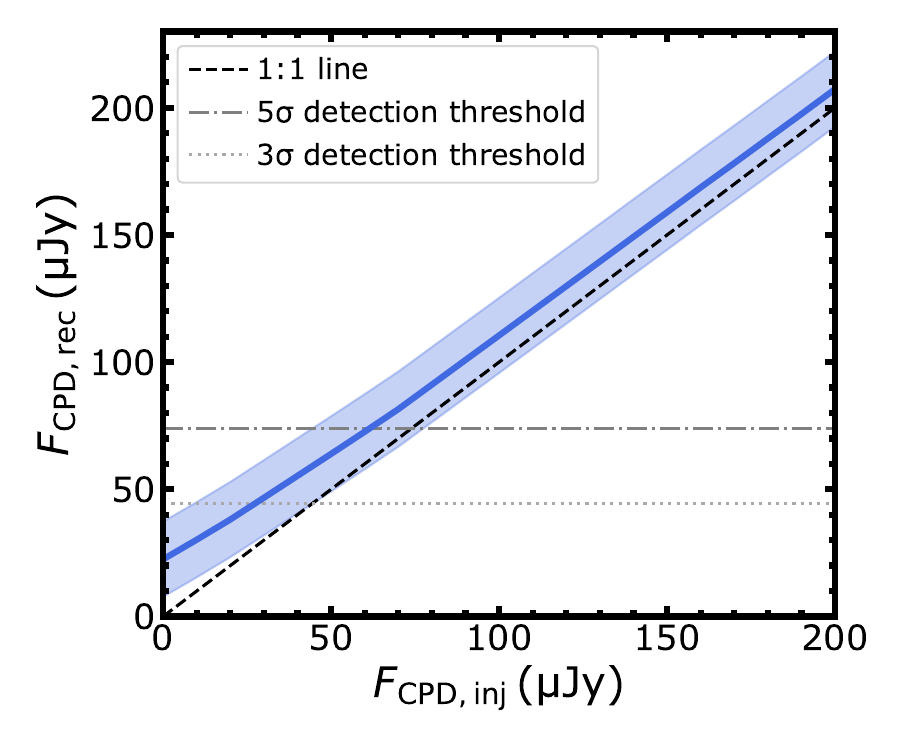}
    \caption{Injected ($F_{\rm CPD,inj}$) vs recovered ($F_{\rm CPD,rec}$) flux density of a point-source at the location of CC1. The legend is the same as for Figure~\ref{fig:CPD_flux_recovery}.}
    \label{fig:CPD_flux_recovery_CC1}
\end{figure}




\end{document}